\documentclass[8.5pt,twoside,twocolumn]{article}
\usepackage[super,sort&compress,comma]{natbib} 
\usepackage[version=3]{mhchem}
\usepackage{balance}
\usepackage{times} %,mathptm}
\usepackage{sectsty}
\usepackage{graphicx} 
\usepackage{lastpage}
\usepackage[format=plain,justification=raggedright,singlelinecheck=false,font=small,labelfont=bf,labelsep=space]{caption} 
\usepackage{fancyhdr}
\begin{document}
  
\thispagestyle{plain}
\fancypagestyle{plain}{
%\fancyhead[L]{\includegraphics[height=8pt]{headers/LH}} %8  
%\fancyhead[C]{\hspace{-1cm}\includegraphics[height=20pt]{headers/CH}}
%\fancyhead[R]{\hspace{10cm}\vspace{-0.25cm}\includegraphics[height=10pt]{headers/RH}}
\renewcommand{\headrulewidth}{1pt}}
\renewcommand{\thefootnote}{\fnsymbol{footnote}}
\renewcommand\footnoterule{\vspace*{1pt}% 
\hrule width 3.4in height 0.4pt \vspace*{5pt}} 
\setcounter{secnumdepth}{5}

\makeatletter 
\renewcommand\@biblabel[1]{#1}            
\renewcommand\@makefntext[1]% 
{\noindent\makebox[0pt][r]{\@thefnmark\,}#1}
\makeatother 
\renewcommand{\figurename}{\small{Fig.}~}
\sectionfont{\large}
\subsectionfont{\normalsize} 

\fancyfoot{}
%\fancyfoot[LO,RE]{\vspace{-7pt}\includegraphics[height=9pt]{headers/LF}}
%\fancyfoot[CO]{\vspace{-7.2pt}\hspace{12.2cm}\includegraphics{headers/RF}}
%\fancyfoot[CE]{\vspace{-7.5pt}\hspace{-13.5cm}\includegraphics{headers/RF}}
\fancyfoot[RO]{\footnotesize{\sffamily{1--\pageref{LastPage} ~\textbar  \hspace{2pt}\thepage}}}
\fancyfoot[LE]{\footnotesize{\sffamily{\thepage~\textbar\hspace{2pt} 1--\pageref{LastPage}}}}
\fancyhead{}
\renewcommand{\headrulewidth}{1pt} 
\renewcommand{\footrulewidth}{1pt}
\setlength{\arrayrulewidth}{1pt}
\setlength{\columnsep}{6.5mm}
\setlength\bibsep{1pt}

\twocolumn[
  \begin{@twocolumnfalse}
\noindent\LARGE{\textbf{Inherent limits on optimization and discovery in physical systems$^\dag$}}
\vspace{0.6cm}

\noindent\large{\textbf{Vladan Mlinar$^{\ast}$\textit{$^{a}$} }}\vspace{0.5cm}
%Please note that \ast indicates the corresponding author(s) but no footnote text is required. 

%\noindent\textit{\small{\textbf{Received Xth XXXXXXXXXX 20XX, Accepted Xth XXXXXXXXX 20XX\newline
%First published on the web Xth XXXXXXXXXX 20XX}}}

%\noindent \textbf{\small{DOI: }}
 \end{@twocolumnfalse} \vspace{0.6cm}

  ]
 
\noindent\textbf{Topological mapping of a large physical system on a graph,  and its decomposition using universal measures is proposed. We find inherent limits to the potential for optimization of a given system and its approximate representations by motifs, and the ability to reconstruct the full system given approximate representations. The approximate representation of the system most suited for optimization may be different from that which most accurately describes the full system.}
\section*{}
\vspace{-1cm}
%Footnotes
\footnotetext{\dag~Electronic Supplementary Information (ESI) available.}
%\footnotetext{\dag~Electronic Supplementary Information (ESI) available. See DOI: 10.1039/b000000x/}

%Please use \dag to cite the ESI in the main text of the article.
%If you article does not have ESI please remove the the \dag symbol from the title and the above footnotetext.

\footnotetext{\textit{$^{a}$~School of Engineering, Brown University, Providence, RI 02912, USA. Tel: +1-401-863-5231; E-mail:vladan$\_$mlinar@brown.edu}}
   
\begin{figure} 
\centering 
\includegraphics[width=0.88\linewidth]{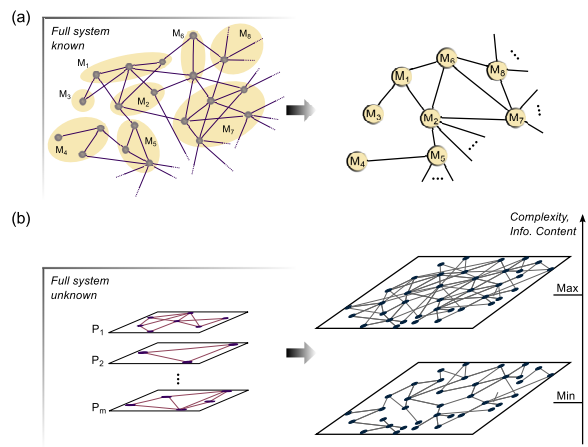} 
\caption{(a) A physical system is represented via a graph where vertices represent distinguishable units. Every interaction between units is depicted by the link connecting units. The representation of the system via graph can be simplified using motifs, denoted as M$_i$, where i = 1, 2, ...,7. (b) Approximate representations by motifs are known, but the full system remains hidden or unknown to us. Complexity measures and information content are used to determine limits of what can be deduced about the full system. ~\label{fig:Fig1a}}                                           
\end{figure}

The analysis of molecular topology in chemistry, where atoms and their connections are represented by a graph, has proven to be a useful means of obtaining the refractive indices, quantum mechanical properties, and biological activity.~\cite{Platt,Trinajstic_2,Bonchev_book,Kermen} We elevate this logic to large physical systems. A large physical system can be represented by a set of distinguishable units which are connected if there is a specific physical interaction between them. In a mathematical framework, each and every unit is replaced by a vertex, and every relationship is replaced by edges connecting corresponding vertices. This creates a graph representation of the system [Fig.~\ref{fig:Fig1a}], which has a certain information content and certain complexity.~\cite{Rashevsky}

The information content of the graph reflects the interactions and grouping between units of the physical system. By adding a new edge (vertex) to the graph, a new interaction (parameter) to the underlying physical system is added, i.e., making it more complex.  Conversely, removing edges and vertices simplifies the system. This means that there is a change in the information content and complexity of the graph, that echoes a change in the physical system. This approach can be used in the analysis of large physical systems, such as large molecules,~\cite{Barone} nanomaterial heterostructures,~\cite{Mlinar_2012} or complex material microstructures,~\cite{Starzewski,Yin_mech} where \emph{ab inito} calculations are prohibitively expensive and/or some underlying physical principles may not be well understood.~\cite{Mlinar_2012,Carter,Tarantola_nature,Mlinar_2013} 

%regardless of the underlying physics of the specific system, 

Starting from graph representations, we use a non-system explicit approach to find inherent limits to optimization and discovery in physical systems. Two challenges naturally arise. \emph{If the full system is known}, there is a need to group smaller parts of the system into lumped conceptual units (``motifs''). It becomes necessary to know how best to choose motifs and how much information is lost in this representation. \emph{If only an approximate representation via motifs and their relationships is known},~\cite{Tarantola_nature,Mlinar_2012,Ceder_2013} it is vital to know how much information can be obtained about the full system.

When moving from the full representation of the system to a partial/approximate one [Fig.~\ref{fig:Fig1a}(a)], we are partitioning the full graph by grouping vertices into sets, i.e., ``motifs,'' guided by the underlying physical processes. Approximate representations of the system use only motifs and interactions between them. An illustrative example can be found in materials science, where, in the full atomistic representation, a material is represented by its constituent atoms and their positions.~\cite{Pettifor,Sanchez,Ceder_2013} By introducing structural motifs, we replace atoms and their bonds by a descriptive quantity, e.g., chemical composition, the type and ratio of atoms in the material.~\cite{Ceder_2013} Whereas knowledge of atoms and their positions guarantees knowledge of chemical composition, knowledge of chemical composition, in principle, is not sufficient to uniquely determine the underlying atomistic structure. For a given chemical composition, there are many possible assignments of atoms 
to each of $N$ lattice sites, so-called ``random realizations,'' with identical composition but distinct physical properties, e.g., optical bandgap.~\cite{Mlinar_2012}

We can calculate the information content of the full system and determine how much information is lost when the full system is represented by motifs using Shannon's information theory.~\cite{Shannon} We define it as:~\cite{Emmert-Streib,Rashevsky,Bonchev_book} $I_{ve}(G) = -\sum_i^{N_m}\sum_j^{N_m}p_{ij}log_2(p_{ij})$, by assigning probabilities to each motif, where $p_{ij}$ is a discrete joint probability distribution that depends on vertices and edges within the motif (Supplementary Information), and $N_m$ is the number of motifs.  

\begin{figure} [h]   
\centering
\includegraphics[width=1.0\linewidth]{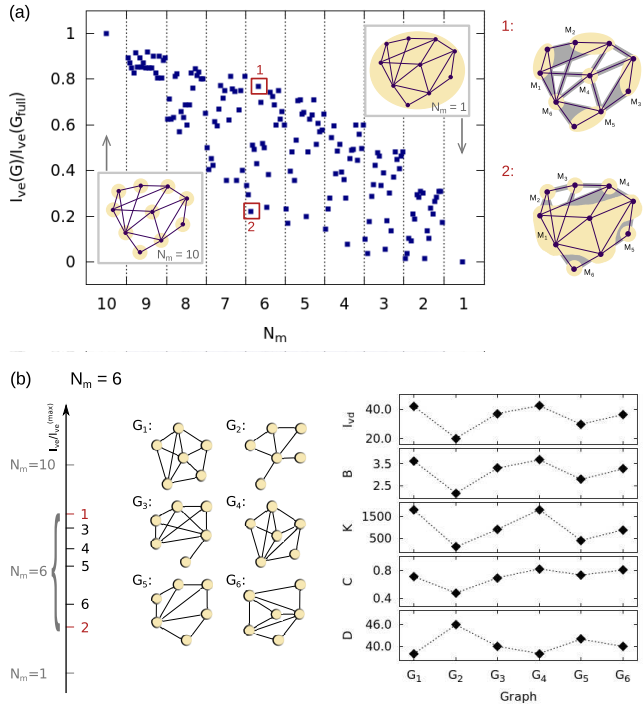}%         
\caption{(a) Information content of the graph normalized to the information content of the full system, I$_{ve}$(G$_{full}$), as it varies with approximate graphs with different numbers of motifs. For $2 < N_m < 9$, there is a vast number of approximate representations of the system, depending on how the motifs are chosen.  A random sample of them are shown by blue squares. Graphs with more vertices and edges aggregated into motifs have less information, e.g., graphs 1 and 2 to right. (b) Characterization of the approximate representations (G$_1$-G$_6$) by the complexity measures, $I_{vd}$, $B$-index, and $K$, the clustering coefficient, $C$, and the graph distance, $D$. \label{fig:Fig2}}          
\end{figure}  

Fig.~\ref{fig:Fig2}(a) shows how $I_{ve}$ varies with $N_m$ for our test case, a system of ten units. Two limiting cases can be identified: (i) each motif contains one vertex, corresponding to the full information content (i.e., no simplification); and (ii) all vertices belong to one motif, corresponding to no information content in the system. For other $N_m$, there are a vast number of approximate representations of the system, depending on how the motifs are grouped. A random sample for $2~<~N_m~<~9$ are shown in Fig.~\ref{fig:Fig2}(a).

The choice of motifs determines how much information is lost relative to the full system. As more vertices and edges are aggregated into a motif, less information is contained in the approximate graph. The choice of motifs is the principal control on information content, surpassing even the number of motifs. This is illustrated in Fig.~\ref{fig:Fig2}(a); for example, many graphs with approximately the same information content can exist for $3~<~N_m~<~7$. Also, Fig.~\ref{fig:Fig2}(a) shows the graphs with the highest [denoted by graph 1 in Fig.~\ref{fig:Fig2}(a)] and lowest [denoted by graph 2 in Fig.~\ref{fig:Fig2}(a)] information content for a fixed value $N_m=6$. 

%Variation in information content could have significant physics-based repercussions. An illustrative example of using approximate representations can be found when analyzing large ($>$10$^4$ atoms) nanostructures, where resolution of structural characterization methods requires introduction of motifs, typically geometry and chemical composition profile. There could be many different approximate representations given the available constraints from the experiment, but all having different physical properties.~\cite{Koenraad}

\begin{figure}   
\centering    
\includegraphics[width=0.85\linewidth]{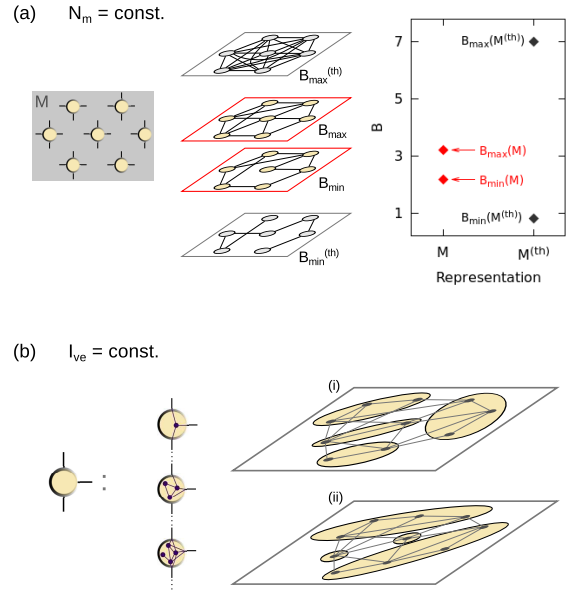}%
\caption{(a) Minimum and maximum complexity (red) of the system based on an approximate graph with seven motifs [Fig.~\ref{fig:Fig2}(a)]. Theoretical mathematical limits are shown in black. (b) Systems with the same numbers of motifs can have different numbers of vertices aggregated in each motif, but the same information content relative to the full system.
\label{fig:Fig3}}      
\end{figure}

Fig.~\ref{fig:Fig2}(b) shows different approximate representations of our test system for $N_m=6$. All of these representations have different information content relative to the full system. However, in order to understand the system as given by an approximate representation, we need to probe the allowed interactions between motifs. In our graph-theoretic approach, this means we need to probe topological features of the graph, and quantify, e.g., connectedness, vertex-vertex separations, etc. This is done by using complexity measures (discussed in detail in the Supplementary Information).~\cite{Rashevsky,Bonchev_book,Platt,Trinajstic_2,Newman,Kermen}

We apply three complexity measures on our six graphs in Fig.~\ref{fig:Fig2}(b): (i) the information content of the vertex degree distribution, $I_{vd}=\sum_{i = 1}^{N_m}a_ilog_2a_i$, where a$_i$ is the vertex degree of a vertex $i$; (ii) complexity index $B$, defined as $B = \sum_{i = 1}^{N_m}a_i/d_i$, where  $a_i$ and $d_i$ are the vertex degree and the distance degree of a vertex $i$, respectively. The vertex distance, $d_i$, is defined by: $ d_i = \sum_{j = 1}^{N_m}d_{ij}$, where $d_{ij}$, the geodesic distance between vertices $i$ and $j$, is defined as the shortest path between them. These distances are calculated using breadth-first search,~\cite{Russell,Newman} and (iii) the total subgraph count, $K$, which includes subgraphs of all sizes, including the graph itself, and is calculated using a recursive algorithm. Each shows that graph G$_4$ exhibits the highest complexity and graph G$_2$, the lowest. We can also analyze how, motifs within given representations, cluster by using the clustering 
coefficient, $C$ (see Supplementary Information).~\cite{Newman,Watts}  Fig.~\ref{fig:Fig2}(b) shows that, whereas G$_3$ has higher complexity than G$_5$ and G$_6$, G$_5$ and G$_6$ have higher clustering coefficients.  Next, the graph distance, $D$ (Supplementary Information), shows that the graph with the highest complexity has the shortest graph distance and vice versa.   

\begin{figure}  
\centering     
\includegraphics[width=0.65\linewidth]{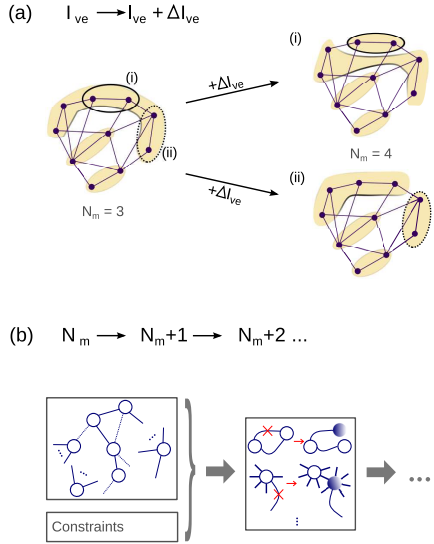}%
\caption{(a) Adding an increment of information, $\Delta I_{vs}$ to the approximate representation for our test system, with N$_m$ = 3: two structures with $N_m+1$ motifs are possible. (b) External constraints should be used to  improve our knowledge of the full system [distinguish between (i) and (ii) in Fig.~\ref{fig:Fig4}(a)].
\label{fig:Fig4}}      
\end{figure}

The role of the complexity measures can be best demonstrated on a practical example of optimization of a second-harmonic-generation device working in resonance conditions.~\cite{Goldoni} The device is a conventional multilayer semiconductor-based structure represented via motifs (materials in each of the layers and barrier, number of layers, size of each individual layer, and chemical composition within the layer). From our perspective, this (approximate) system can be readily represented by a graph. Each motif is represented as a vertex, and depending on the interaction between the motifs, that vertex can or cannot be connected with other vertices. For example, whereas there is a connection between the size, and the material and its composition in that layer, there is no connection, in this representation, between the size of a layer and the material in the barrier. Knowing the allowed connections between vertices, we can calculate the complexity measures and determine their minimum and maximum values for 
this approximate representation. If we want to understand, for example, how interfaces between layers influence the behavior of the system, we need to generate a new approximate system whose complexity is higher than the maximum complexity of the previous system. This shows that the initial approximate system is not sufficient to describe a given effect. This type of analysis could enable us to find optimal representations of physical systems via motifs and form the basis for the automatic design of motifs for targeted applications.

Next issue is how much of the full system can be reconstructed if only partial views (approximate representations) of the system are known [Fig.~\ref{fig:Fig1a}(b)]. This would mean probing all the available relations between motifs and disaggregating the motifs.  Even in our test case [Fig.~\ref{fig:Fig2}(a)], it is difficult to imagine how to reconstruct the full system containing ten vertices from one of the approximate representations in Fig.~\ref{fig:Fig2}(b). However, this is not an uncommon situation in physics. One example is the dissipation of energy in solid materials undergoing cyclical deformation.~\cite{Jackson_2002} The dependencies of dissipation on easily measured physical motifs such as grain size may be controlled by a number of underlying physical properties such as volume and grain boundary diffusivity and grain boundary thickness, etc. However, the underlying physical processes controlling dissipation, and the length scales over which they operate, are not always well understood.~\cite{
Jackson_2002} If we are able to find inherent limits to optimization and discovery in physical systems, we can then explore options for development of automatized procedures in discovery of new systems.  

For a given approximate representation and existing physical constraints on motifs, we can determine the limiting cases of minimally interacting and fully interacting motifs. Within our graph-theoretic approach, this means investigating allowed connections between vertices and calculating the minimum and maximum complexity of the approximate representation. This is illustrated in Fig.~\ref{fig:Fig3}(a) for our test case using an approximate graph with $N_m$ = 7. Physical constraints (specific characteristics of the system) determine which connections are allowed. Without the constraints, the minimum and maximum complexity have no physical meaning and are purely mathematical limits: The maximum corresponds to the complexity of the complete graph with seven nodes, and the minimum corresponds to a graph where no node is disconnected. The minimum and maximum complexity values arising from a system with physical constraints necessarily lie in between the mathematical limits. This picture gives inherent limits for 
optimization of the given approximate representation. However, it does not give any new insights into the full system. Fig.~\ref{fig:Fig2}(b) illustrates this for our test case: the approximate representation of the system most suited for optimization (graph G$_4$) may be different from that which has the highest information content, or equivalently, most accurately describes the full system (graph G$_1$).

To reconstruct/discover an unknown full system starting from a given approximate representation, we need to disaggregate the motifs to determine what vertices and edges might be inside. This could be very difficult because, for a representation with the same number of motifs and even the fixed information content, the number of vertices contained in each motif may vary. Fig.~\ref{fig:Fig3}(b) illustrates this for our test system [Fig.~\ref{fig:Fig2}(a)]; both approximate representations have the same number of motifs and the same information content relative to the full system.

The introduction of additional information can disaggregate motifs. For example, Fig.~\ref{fig:Fig4}(a) shows that in an approximate system with $N_m=3$, adding an increment of information, $\Delta I_{vs}$, gives an approximate system with $N_m+1$ motifs, improving our knowledge of the system. However, Fig.~\ref{fig:Fig4}(a) also shows that the same starting point and $\Delta I_{vs}$, gives two different \emph{possible} systems (scenarios (i) and (ii)). Of course, databases and external constraints imposed for a specific problem can improve our knowledge of the full system [Fig.~\ref{fig:Fig4}(b)], e.g., to distinguish between (i) and (ii) in Fig.~\ref{fig:Fig4}(a). In this sense, this type of logic is useful in cases when we can guide the solution using external constraints or previous knowledge, for example, to automatically search for motifs with targeted applications in mind. However, this may not be possible for understanding a fully unknown system, where the appropriate semantic framework is absent.

%Also, by extending this logic, we can regard the initial system with ten units and each motif containing one unit in Fig.~\ref{fig:Fig2}(a) as representing the maximum of our current knowledge. If additional information is added to the system, $N_m \rightarrow N_m+i$, where i$\geq$1, [Fig.~\ref{fig:Fig4}(b)], then our initial system becomes just one of many possible representations for $N_m$ = 10, of the new full system. However, this type of analysis may not be possible for understanding a fully unknown system, where the appropriate semantic framework is absent.

In summary, we applied graph theoretic approach and universal measures to analyze full and partial representations of physical systems. We find inherent limits to the potential for optimization of a system represented by motifs, and the ability to reconstruct the full system given approximate representations. The approximate representation most suited for optimization may be different from that which most accurately describes the full system. This approach could enable us to find optimal representations of physical systems via motifs and form the basis for the automatic design of motifs for targeted applications. 

This work was supported by NASA EPSCoR. We thank A.C. Barr for reading of the manuscript. 

\footnotesize{
\bibliography{inhLim_VMlinar} 

\newpage

\noindent \textbf{ELECTRONIC SUPPLEMENTARY INFORMATION} \\ \\ %\vspace{1.6cm}
%  \end{@twocolumnfalse}
% 
%   \begin{@twocolumnfalse}
 \noindent\Large{\textbf{Inherent limits on optimization and discovery in physical systems}} \\ \\  \\
 %\vspace{0.2cm}
% 
 \noindent\large{\textbf{Vladan Mlinar$^{\ast}$} \vspace{0.5cm}}
 
\renewcommand\thefigure{S.~\arabic{figure}}  
\renewcommand{\thesection}{S.~\Roman{section}}
\renewcommand\theequation{S.~\arabic{equation}}

\setcounter{page}{1}
\setcounter{figure}{0} 

\section*{\large{Complexity, information content, and complexity measures/indices}}
\footnotetext{\textit{School of Engineering, Brown University, Providence, RI 02912, USA;  Tel: (+1)401-863-5231; E-mail: vladan$\_$mlinar@brown.edu}} 

\normalsize

In the framework of the information theory, an arbitrary system, under certain conditions can be analyzed in terms of its information content and complexity.~\cite{Rashevsky,Newman,Trinajstic_2,Bonchev_book,Kermen,Hendrickson} Complexity and the information content are directly proportional, i.e., in order to have a sufficiently large information content, the system must be sufficiently complex.~\cite{Rashevsky} This type of analysis has been applied to various systems ranging from cellular and ecological networks to chemical structures.~\cite{Trinajstic_2,Kermen} For example, as already discussed in the main body of the manuscript, different arrangements of atoms in a molecule can be represented by the graph, and by analyzing topological features of the graph, i.e., the molecular topology, one can determine refractive indices, quantum-mechanical properties and biological activities of molecules.~\cite{Trinajstic_2} Also, one can count the subgraphs containing two edges (two-bond molecular fragments), known 
as Platt's index in chemical theory.~\cite{Platt,Bonchev_book}

% 
% The system gets represented via different and distinguishable units, in the form of a graph with vertices and edges, and the information content and complexity can be, to the large extent, measured and characterized through the graph's topology. Various quantitative measures, often called indices, have been in use, typically information-theoretical measures, based on Shannon's information~\cite{Shannon}, or topological complexity measures that e.g., analyze vertex-vertex connectedness and vertex-vertex separations, count the simple subgraphs with targeted number of edges, etc.~\cite{Bonchev_book,Kermen}  %This suggests that one can put a lower limit on the complexity of a system.
% 
% After a physical system gets mapped to a graph (basic notations and graph descriptors are presented in Sec.~\ref{sec:basicFeatures}), we can analyze the graph in terms of its information content and complexity~\cite{Rashevsky,Newman,Trinajstic_2,Bonchev_book,Kermen,Trinajstic,Bertz,Hendrickson}. Complexity and the information content are directly proportional, i.e., in order to have a sufficiently large information content, the system must be sufficiently complex~\cite{Rashevsky,Bonchev_book}. 

Prior to discussing various complexity measures, for the clarity reasons and the completeness of the work, we briefly review basic definitions of the graphs, including the basic graph descriptors (Sec.~\ref{sec:basicFeatures}).

There is a vast number of complexity measures and indices proposed,~\cite{Bonchev_book,Binder} some of them have been defined with a specific application in mind, other to cover certain aspects of the structure of the graph, etc. Of course, the goal of this work is not to propose a yet another complexity measure, but to discuss new concepts using well-known and well-established complexity measures. In the following sections we will describe widely-used measures/indices of  the topological complexity (Sec.~\ref{sec:topCompl}), and measures/indices of the compositional complexity, which exploits the equivalence and diversity of the elements of the studied system (Sec.~\ref{sec:compCompl}). The former is relevant for our concepts of studying optimization and design issues for a given set of approximate representations of a system, whereas the latter is relevant for our concept of introducing "motifs" and tracking the information loss from full to partial representations of the system.

% For completeness, we mention here the algorithmic information, as defined by Kolmogorov~\cite{Kolmogorov}. The corresponding measure of complexity is defined as the shortest computer program capable of generating a given string~\cite{Kolmogorov,Bonchev_book,Binder}. 

% \section{Structural analysis of graphs} 
% 
% After a physical system gets mapped to a graph, the original system is no longer used in further analysis. The problem arising from this is that all the relationships that existed in the system and/or between the two real-world systems need to be compared by measures between the abstract constructs such are graphs (representing the original objects). We need to be able to decide the relationship between these graphs. Whereas in the real world one could always revere to the expert knowledge of the observer, in our formal framework the relations are probed by a mathematical measure. In general, we view the mathematical  measure as a meaningful if it corresponds to the measure obtained in the real-world, from expert knowledge. 
% 
% In what follows, we define the basic graph descriptors (Sec.~\ref{sec:basicFeatures}), complexity measures (Sec.~\ref{sec:Complexity}), and information-theoretical measures in the case of hierarchical graphs (Sec.~\ref{sec:infotheory}). 

\section{Basic graph descriptors}
\label{sec:basicFeatures}

Graphs are represented by the set of $n$ vertices and the set of $m$ edges, where the edge $\{ij\}$ is the line that emanates from vertex $i$ and ends in vertex $j$. In this work we are focused on the simple, connected, undirected graphs. The ``simple graph'' is a graph with \emph{no} multiple edges, i.e., pairs of vertices linked by more than one edge and \emph{no} loops, i.e., edges that begin and end at the same vertex. A graph is connected if there is a path between any pair of vertices in it and undirected if all of its edges are undirected. %If any two vertices are connected by an edge, a graph is complete. %For more formal mathematical definitions in graph theory see e.g.,    

Two vertices $j$ and $i$ are called adjacent if they are connected by an edge $\{ij\}$.  The adjacency relation is quantified by the term $a_{ij} = 1$ and the no adjacency by $a_{ij} = 0$. The number of the nearest-neighbors of a vertex $i$ is termed the \emph{vertex degree}, $a_i$, and is given by 
\begin{equation}
a_i = \sum_{j = 1}^{n}a_{ij}~, ~\label{eq:ai}
\end{equation}
where $n$ is the total number of vertices in the graph $G$. $a_i$ is one one of the local connectivity descriptors.

The sum of all vertex degrees in a graph $G$ defines its \emph{total adjacency}, $A(G)$:
\begin{equation}
A(G) = \sum_{i = 1}^{n}\sum_{j = 1}^{n}a_{ij} = \sum_{i = 1}^{n}a_i~.
\end{equation}

Next, the \emph{average vertex degree} is defined by:
\begin{equation}
 \langle a_i\rangle = \frac{A}{n}
\end{equation}

The \emph{connectedness} is defined by:
\begin{equation}
 Conn = \frac{A}{n^2} = \frac{2m}{n^2}~,
\end{equation}
where $m$ is the total number of edges in the graph.

One can also define connectedness by:
\begin{equation}
 Conn' = \frac{A}{n(n-1)} = \frac{2m}{n(n-1)}~.
\end{equation}
Here $n(n-1)$ is the number of edges in a complete graph (If any two vertices are connected by an edge, a graph is complete).

The \emph{clustering coefficient} is introduced to describe the property of transitivity of graphs. Namely, it is found in many graphs that if vertex A is connected to vertex B and vertex B to vertex C, then there is a heightened probability that vertex A will also be connected to vertex C.~\cite{Newman} Thus, one can define a local value of the cluster coefficient for each vertex, $c_i$, and is defined by:~\cite{Newman}
\begin{equation}
 c_i = \frac{2m_i}{a_i(a_i-1)}~, \label{eq:Ci}
\end{equation}
where $m_i$ is the number of edges between the first neighbors of the vertex $i$ and $a_i(a_i-1)/2$ is $m_i(max)$ (for the complete graph). For vertices with degree 0 or 1, for which both numerator and denominator are zero, we put $c_i$ = 0. Note that, in addition to the vertex degree [Eq.~(\ref{eq:ai})], the local clustering coefficient is another \emph{local connectivity descriptor}. 

The clustering coefficient for the whole graph/network is the average over vertices of the local clustering coefficients:~\cite{Newman,Watts}
\begin{equation}
 C = \frac{1}{n}\sum_ic_i~. \label{eq:C}
\end{equation}

Globally, the \emph {$n^{th}$ extended connectivity}, $^nEC$ is introduced, which takes into account the layers of first-, second-, third-, etc. neighbors. It is defined as: 
\begin{equation}
 ^nEC = \sum_{i = 1}^na_i = \sum_{i = 1}^n\sum_{j adj i}^{n-1}a_j~. \label{eq:EC}
\end{equation}

Similarly to the adjacency for vertices and the full graph, one can define distances. The \emph{(geodesic) distance}, $d_{ij}$ between vertices $i$ and $j$ is defined as the shortest path between them. These distances are calculated using breadth-first search.~\cite{Russell,Newman} The \emph{distance degree}, or vertex distance, $d_i$ is defined by: 
\begin{equation}
 d_i = \sum_{j = 1}^nd_{ij}
\end{equation}

The \emph{graph distance}, $D$, is defined as 
\begin{equation}
 D(G) = \sum_{i = 1}^n\sum_{j = 1}^n d_{ij} = \sum_{i = 1}^{n}d_i \label{eq:D}
\end{equation}

The \emph{average vertex distance (degree)}, $\langle d_i \rangle$, is defined by:
\begin{equation}
 \langle d_i \rangle = \frac{D}{n}
\end{equation}

The \emph{average graph distance} (also called the graph radius, or average path length, or average degree of vertex-vertex separation) is given by
\begin{equation}
 \langle d \rangle = \frac{D}{n(n-1)}
\end{equation}

For an undirected graphs, one can introduce the \emph{mean geodesic} which is the shortest distance between vertex pairs in a graph, $l$~\cite{Newman,Watts}:
\begin{equation}
 l = \frac{1}{\frac{1}{2}n(n+1)}\sum_{i\geq j}d_{ij}~. \label{eq:l}
\end{equation}

One needs to be careful when judging the importance and potential of these descriptors. For example, the mean geodesic $l$ was shown to be quite relevant in analyzing ``small world'' networks, in particular in studies of disease spreading.~\cite{Watts}

\section{Topological complexity}
\label{sec:topCompl}

%\subsection{Complexity measures/indices}
%\label{sec:Complexity}   

As we have already mentioned, there is no single measure of the complexity of a graph. There is a vast number of different measures throughout the literature, all capturing some aspects of the graphs' complexities, some very similar, etc.   
Here we present the measures that have been relevant to the current work:

-- \emph{The information content of the vertex degree distribution of a graph} is defined using Shannon's theory,~\cite{Shannon} and is given by:~\cite{Bonchev_1}

\begin{equation}
 I_{vd}=\sum_{i = 1}^{n}a_ilog_2a_i~, \label{eq:Ivd}
\end{equation}
where $a_i$ is the vertex degree (Eq.~(\ref{eq:ai})). I$_{vd}$ increases with the connectivity and other complexity factors such as the number of branches, cycles, cliques, etc.

-- \emph{The global edge complexity} is actually the total adjacency, A, and is defined as: 

\begin{equation}
 E_g = A = \sum_i^n\sum_i^na_{ij} = \sum_i^na_i  \label{eq:Eg}
\end{equation}

-- \emph{The average edge complexity} is actually the average vertex degree, $\langle a_i\rangle$, and is defined as: 

\begin{equation}
 E_a = \langle a_i\rangle = \frac{A}{n} = \frac{E_g}{n}~, \label{eq:Ea}
\end{equation}

-- \emph{The normalized edge complexity}, E$_n$, is actually the connectedness,

\begin{equation}
 E_n = Conn = \frac{A}{n^2} = \frac{E_g}{n^2} \label{eq:Conn}
\end{equation}

-- \emph{The subgraph count indices} are complexity measures based on counting the simple subgraphs with a given number of edges. The second order subgraph-count index counts all subgraphs with two edges and is denoted as $^2SC$. The normalized second order sugraph count is given by:~\cite{Bonchev_book,Platt}
\begin{equation}
^2SC_n =\frac{^2SC}{^2SC(K_v)} = \frac{^2SC}{0.5V(n-1)(n-2)} \label{eq:SCn}
\end{equation}

This can be further expanded to lead to the \emph{total subgraph count}, $K$, which includes subgraphs of all sizes, including the graph itself, regarded as a proper subgraph. We developed an in-house, recursive algorithm to count all subgraphs with a given number of edges. It also allow us to calculate the overall connectivity (see next indices). 

-- \emph{The overall connectivity} represents a set of indices which defines a certain overall graph-invariant X, by the sum over the values this invariant has for each of the subgraphs. Also, the contributions of all subgraphs having k edges are combined in a single term, $^kX$. A typical example of this approach is when X is chosen to be the graph adjacency. Then, the overall connectivity, as a complexity measure is defined as:

\begin{equation}
 OC(G) = \sum_{k=1}^m{ ^k}OC = \sum_{k=1}^m \sum_i{ ^k}A_i(^kG_i \subset G) \label{eq:OC}
\end{equation}

\begin{equation}
 \{OC\} = \{{^0}OC, {^1}OC, {^2}OC,...,{^E}OC\} 
\end{equation}

- \emph{The A/D index} is a complexity measure that focuses on vertex-vertex connectedness and vertex-vertex separation of a graph. Namely, it was shown that graphs with high complexity are characterized by both high vertex-vertex connectedness and small vertex-vertex separation (the small world concept of Watts and Strogatz).~\cite{Watts} Thus, we can use both quantities in characterizing network complexity. The A/D index is defined as follows:~\cite{Bonchev_book}

\begin{equation}
 A/D = \frac{\langle a_i\rangle}{\langle d_i \rangle}~, \label{eq:AD}
\end{equation}
where $\langle a_i\rangle$ is the average vertex degree and $\langle d_i \rangle$ is the average distance degree. At a constant number of vertices, the A/D index has a minimum value in path graphs (characterized by low connectivity and long distances), and has maximum value in the complete graphs (which are maximally connected and all of their vertices have only a unit distance separation.

- \emph{The complexity index B} is defined as the sum over the $b_i$ values of all graph vertices, where the ratio $b_i = a_i/d_i$ of the vertex degree $a_i$ and its distance degree $d_i$ is a local invariant with interesting centric properties:~\cite{Bonchev_book} 

\begin{equation}
 B = \sum_{i = 1}^{n}\frac{a_i}{d_i} \label{eq:B}
\end{equation}

Note that \emph{B index} is expected to behave similarly to the \emph{A/D index}, with less degeneracy, and more sensitivity to local topology. 

\section{Basic graph descriptors and complexity measures for graph representations in Fig. 2(b)}

Fig.~\ref{fig:S1} shows how the basic graph descriptors and complexity measures are used for characterization of the approximate graph representations G$_1$--G$_6$ (see also the main body of the manuscript and Fig.~2). The basic graph descriptors of the graphs are defined in Sec.~\ref{sec:basicFeatures}, and complexity measures in Sec.~\ref{sec:topCompl}.

\begin{figure}[h]  
\centering
\includegraphics[width=1.00\linewidth]{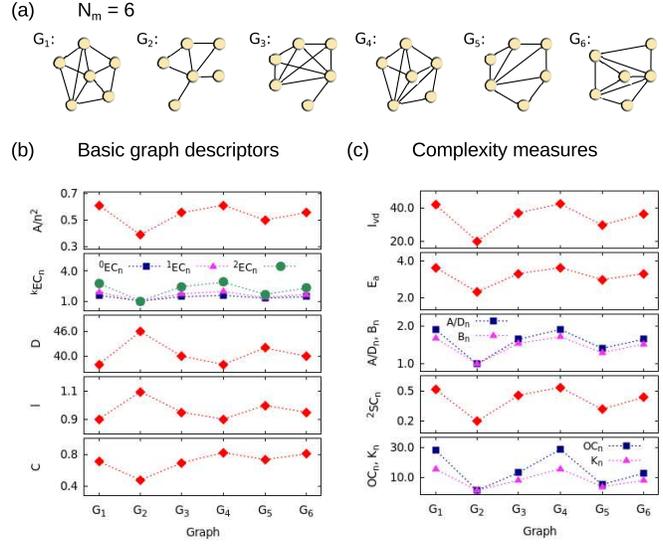}%
\caption{(a) Approximate graph representations of a system discussed in Fig.~2(b). (b) Basic graph descriptors of the graphs (defined in Sec.~\ref{sec:basicFeatures}). Note that the extended connectivities are normalized on the minimum value for the graphs (in this case graph $G_2$), $^0EC_{n} = ^0EC/^0EC_{min}$, $^1EC_{n} = ^1EC/^1EC_{min}$, $^2EC_{n} = ^2EC/^2EC_{min}$; and (c) complexity measures (defined in Sec.~\ref{sec:topCompl}). A/D index, the complexity index, B, the subgraph count index, $K$, and overall connectivity index, $OC$, are normalized on the minimum their corresponding minimum values for the graphs (in this case graph $G_2$). ~\label{fig:S1}}
\end{figure}

Note that these descriptors and measures exhibit different sensitivity to the local topology of graphs. This is best illustrated by looking at the difference between basic descriptors and complexity measures for the graphs G$_1$ and G$_4$. \emph{For basic descriptors}, A/n$^2$, $EC(0)$, $D$, and $l$ do not distinguish between the graphs G$_1$ and G$_4$, but $EC(1)$, $EC(2)$, and $C$ do. \emph{For complexity measures}, $E_a$, $A/D$ find that the graphs G$_1$ and G$_4$ have the same complexity, whereas the measures $I_{vd}$, $B$, $^2SC_n$, $K$, and $OC$ find that the graph G$_4$ is more complex than G$_1$. The reason for the discrepancy between the different measures lies in their sensitivity (or lack of it) to the local topology of the graph. Subtle variations in local graph topology which are not captured by a given measure, e.g., A/D index are captured by B-index, K, or OC.

\section{Compositional complexity}
\label{sec:compCompl}

It was discussed in the literature that measures of structural or topological complexity should not be based on symmetry, because symmetry is a simplifying factor.~\cite{Rashevsky,Emmert-Streib, Kermen,Bonchev_book} However, use of symmetry is justifiable in defining the compositional complexity, which is based on equivalence and diversity of the elements of the system studied. Traditionally, vertices of an undirected and unweighted graphs are partitioned in sets of indistinguishable vertices according to their dependence on local and non-local degree-dependences.~\cite{Rashevsky} However, in the case where there is a preexisting partition of the graph, one can define levels containing pre-defined vertices and edges.~\cite{Emmert-Streib}

\subsection{An information-theoretic approach}
\label{sec:infotheory}

As discussed in the main body of the manuscript, we can calculate the information content of the full system and determine how much information is lost when the full system is represented by motifs using Shannon's information theory.~\cite{Shannon} 

We define it as:~\cite{Emmert-Streib,Rashevsky,Bonchev_book} $I_{ve}(G) = -\sum_i^{N_m}\sum_j^{N_m}p_{ij}log_2(p_{ij})$, by assigning probabilities to each motif, where $p_{ij}$ is a discrete joint probability distribution that depends on vertices and edges within the motif, and $N_m$ is the number of motifs. 

$p_{ij}$ is given by: 

\begin{equation}
 p_{ij} = \left\{
 \begin{array}{rl}
  \frac{p_i^vp_j^e}{\sum_jp_j^vp_j^e}, & i = j,\\
  0, & i \neq j.
 \end{array} \right.~\label{eq:pij}
\end{equation}

In Eq.~(\ref{eq:pij}), $p_i^v = n_i/N_v$ and  $p_i^e = e_i/(2N_e)$, where $n_i$ and $n_e$ are the number of vertices, and the number of edges of all vertices in motif $i$, respectively. $N_v$ and $N_e$ are the total number of vertices and the total number of edges in the given graph representation, respectively.

\section{An example:  Full structure versus approximate representations by motifs}

The structure is represented via atoms and their positions. This gives the full structural information about the system, including a specific physical property (Fig.~\ref{fig:struct_prop}). For several-atom systems, the geometrically possible structures and chemical bonds are well understood. Thus, it has become possible to predict new materials on this scale that have targeted physical/mechanical properties (the field of materials informatics).~\cite{Deaven,Ortiz}

However, the structure of larger systems is typically represented via descriptive quantities, ``structural motifs,'' such as composition profile and geometry ($ii$ in Fig.~\ref{fig:struct_prop}),~\cite{Mlinar_2012} confining potential ($iii$ in Fig.~\ref{fig:struct_prop}),~\cite{Goldoni} or representative volume elements (RVEs) ($iv$ in Fig.~\ref{fig:struct_prop}). For example, continuum mechanics uses the concept of RVE~\cite{Starzewski,Yin_mech}, to approximate the true material. An RVE needs to be spatially invariant and large enough to contain a sufficient number of micro-features to represent the entire material microstructure.~\cite{Starzewski} When a material is parametrized, the parameters are defined to quantitatively describe attributes such as particles, voids, grains, and micro-cracks.~\cite{Starzewski,Yin_mech}

\begin{figure} [h]
\centering
\includegraphics[width=1.0\linewidth]{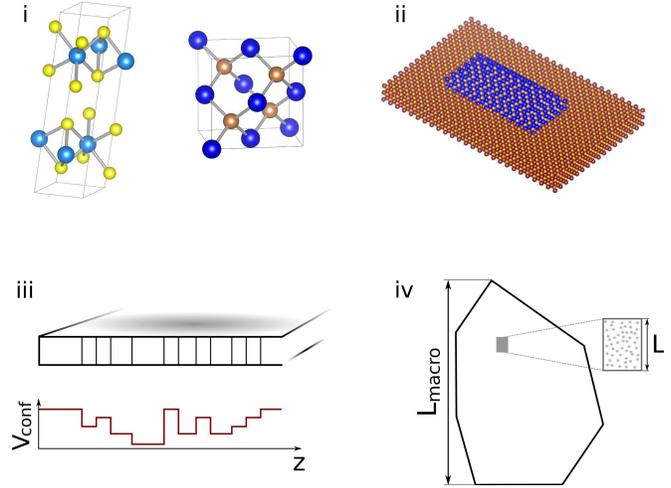}%
\caption{Description of the structure: (i) on the atomistic level, examples of  MoS$_2$ and InAs; (ii) on the semi-empirical atomistic level ($> 10^4$ atoms), where, typically, structural motifs such as the chemical composition and geometry are introduced; example of InSb/GaAs quantum dots; (iii) on the mesoscopic level, where the confining potential is used; example of superlattice; and (iv) on the continuum level, using the concept of a representative volume element (RVE). An RVE plays the role of a mathematical point of continuum field approximating the true material. Note the separation of scales: the microscale $d$, such as e.g., average size of grain in a given microstructure, the mesoscale $L$ (the size of the RVE), and the macroscale L$_{macro}$. \label{fig:struct_prop}}
\end{figure}

Describing the structure via motifs leads to the loss of structural information, and knowledge of the full structural information is replaced by the partial structural information contained in motifs. For example, at scales of even only up to a few nanometers, atomic-scale effects are sometimes overlooked, depending on how the structure is described; a simple example is the effect of atomic scale randomness in alloys. For a given chemical composition of an alloy, there are many different assignment of atoms to each of the $N$ lattice sites, so called ``random realizations,'' that have the same composition but distinct physical properties, e.g., optical bandgap. Depending on how the structure is described, or to be precise, parametrized in a model, the effects on the atomistic level, including the effect of atomic-scale randomness, will or will not be included (see $ii$ and $iii$ in Fig.~\ref{fig:struct_prop}). If a structure is described only via chemical composition and geometry, one will not be able to  distinguish between different random realizations.~\cite{Mlinar_2012}

%\newpage
%\footnotesize{
%\bibliography{inhLim_VMlinar} %your .bib file
%\bibliographystyle{rsc} %the RSC's .bst file
%}

\end{document}